\documentstyle[aps,pra,twocolumn,amssymb,eqsecnum]{revtex}
\newcommand{\bea}{\begin{eqnarray}}
\newcommand{\eea}{\end{eqnarray}}
\newcommand{\beq}{\begin{equation}}
\newcommand{\eeq}{\end{equation}}
\newcommand{\cam}{{\cal M}}
\newcommand{\pa}{\partial}
\newcommand{\snp}{{\sum_N}'}
\newcommand{\ina}{\frac 1 {2\pi i} \int\limits_{c-i\infty}^{c+i\infty}}
\newcommand{\ing}{\ina d\alpha \,\,\Gamma (\alpha)}
\newcommand{\en}{\left( E_N -\mu_c \right)}
\newcommand{\reals}{\mbox{${\rm I\!R }$}}
\newcommand{\rez}{\mbox{ Res } \zeta }
\newcommand{\nn}{\nonumber}
\newcommand{\nats}{\mbox{${\mathbb N }$}}
\newcommand{\intgs}{\mbox{${\mathbb Z }$}}
\newcommand{\muc}{\left(e^{-\beta (\mu_c -\mu)}\right)}

\begin{document}

\title{Bose-Einstein condensation in arbitrarily shaped cavities}  

\author{Klaus~Kirsten    \thanks{E-mail address:
{\tt kirsten@itp.uni-leipzig.de}}}
\address{University of Leipzig, Institute of Theoretical Physics\\
Augustusplatz 10, 
04109 Leipzig} 
\author{David J. Toms   \thanks{E-mail address:
{\tt d.j.toms@newcastle.ac.uk}}}
\address{Department of Physics, University of Newcastle Upon Tyne,\\
Newcastle Upon Tyne, United Kingdom NE1 7RU}

\date{\today}
\maketitle
\begin{abstract}
We discuss the phenomenon of Bose-Einstein condensation of an ideal
non-relativistic Bose gas in an arbitrarily
shaped cavity. The influence of the finite extension of the cavity on
all thermodynamical quantities, especially on the critical temperature
of the system, is considered. We use two main methods which are 
shown to be equivalent. The first deals with the partition 
function as a sum over energy levels and uses a Mellin-Barnes 
integral representation to extract an asymptotic 
formula. The second 
method converts the sum over the energy levels to an integral with a 
suitable density of states factor obtained from spectral analysis. 
The application to some simple cavities is discussed.
\end{abstract}

\section{Introduction}
It is now well over seventy years since the phenomenon referred to as
Bose-Einstein condensation (BEC) was first predicted \cite{1,2}.
Until recently the best experimental evidence that BEC could occur
in a real physical system was liquid helium, as suggested originally
by London \cite{3}. More recently it was suggested \cite{4,5} that
BEC could occur for excitons in certain types of non-metallic 
crystals (such as
CuCl for example). There is now good evidence for this in a number of
experiments \cite{6}. However, the most exciting experimental evidence
for BEC has come from the observations of very cold alkali gases.
BEC has now been observed to occur in gases of rubidium \cite{7}, 
lithium \cite{8} and sodium \cite{9}. In these experiments BEC has been
realized in magnetic traps and laser traps, in small volumes far away
from the infinite volume limit. 
Although nowadays a few million particles remain in the trap, 
in these early experiments only up to 2000 particles
remained there so that it was by no means clear if calculations done in
the thermodynamical limit are an accurate treatment when 
considering this experimental situation. (For a more detailed analysis of this see Refs.~\cite{GHPLA,GHNat,bros,DJTKK,ravndal,stef}.) This
motivates us to consider the related problem of the quantum statistics of a finite
number of particles in an arbitrarily shaped cavity.

When thinking about quantum statistics in finite volumes there appears
an immediate difference to the thermodynamic limit. Quantum particles 
confined to a finite volume of arbitrary size inevitably have a nonzero 
kinetic energy, hence they must exert some pressure. Taking 
$\mu m$-sized cavities and the parameters typical for liquid helium it 
can be seen that the zero-point pressure is of measurable magnitude 
\cite{10}. But not only the pressure but also all other thermodynamical 
quantities may deviate substantially from the results obtained in the 
thermodynamical limit once the volume and the particle number get small
enough. In this article we want to give a systematic treatment of this
problem.

In doing this we are going to extend a recent careful and beautiful study
of BEC in a cubic volume where the gas was supposed to fulfill Dirichlet
boundary conditions at the boundary of the cavity \cite{10}. The authors 
there employed an approach where the sum over the discrete spectrum was
replaced by an integral with an appropriate density of states. This density
was obtained 
by the knowledge of the energy spectrum in this example.

In this article we are going to treat an arbitrarily shaped cavity. 
The methods we are going to employ are the so called heat-kernel 
techniques extensively used in finite temperature relativistic quantum
field theory starting with 
the work of Dowker, Kennedy \cite{11}. However, in these
considerations the stress was more on the influence of gravitational 
fields on the quantum statistics (see also \cite{12}). Recently also the
influence of boundaries was considered in the context of the Casimir
effect \cite{13,14} and of BEC as symmetry breaking \cite{15,16}. In
non-relativistic theories these techniques are however nearly unemployed
and we want to show in the present article that they also can be used here in a very effective way. 

The plan of the paper is as follows. First we develop the quantum statistics
of $N$ noninteracting bosons in a finite cavity of arbitrary shape.
We will exemplify the use of heat-kernel techniques and of the Mellin-Barnes
integral representation for the calculation of the partition function. We
will explicitly show what influence the boundary and its shape have on 
all thermodynamical quantities, especially on the critical temperature
at which BEC occurs.
In Sec.~III we describe an alternative approach where sums are converted into
ordinary integrals with an appropriate density of states factor. We will 
show that the density of states is determined by the heat-kernel 
coefficients of the associated Schr\"odinger operator. The next section 
is devoted to the discussion of some specific examples like the rectangular 
box and the spherical cavity with various boundary conditions. In the 
final section we present a short summary of our main results.

\section{Quantum statistics of a free Bose gas in a 
d-dimensional finite cavity}
Let us consider a system of $N$ noninteracting bosons in a finite cavity 
$\cam$ with boundary $\pa \cam$ and with single particle state energies 
$E_N$ determined by the Schr\"odinger equation
\beq
-\frac{\hbar ^2 }{2m} \Delta \phi_N (x) =E_N \phi_N (x) ; 
\left.  \qquad \phi_N \right|_{x\in\pa \cam}=\ldots, \label{1}
\eeq
where we do not need to fix here the boundary conditions imposed on the 
field because our treatment will be quite general.

In the grand canonical approach, the partition sum then reads
\beq
q=-\sum_N \ln \left( 1-z e^{-\beta E_N }\right), \label{2}
\eeq
with the fugacity $z=\exp (\beta \mu )$, $\mu$ being the chemical potential
and $\beta = 1/(kT)$. In the discussion of BEC the ground state always 
plays a special role and for that reason we write
\beq
q=q_0 -  { \sum_N}' \ln \left( 1-z e^{-\beta E_N }\right). \label{3}
\eeq
Here the prime indicates that the ground state is to be omitted and $q_0$ 
is the contribution of the ground state with energy $E_0$ and degeneracy
$d_0$,
\beq
q_0 = -d_0  \ln \left( 1-z e^{-\beta E_0 }\right). \label{4}
\eeq
For the calculation of the partition sum, equation (\ref{3}), we will first
expand the logarithm to obtain
\beq
q=q_0 +\sum_{n=1}^{\infty} \snp \frac 1 n e^{-\beta n (E_n -\mu) }.\label{5}
\eeq
For the evaluation of this kind of expressions it is very effective to
make use of the Mellin-Barnes type integral representations,
\beq
e^{-v} = \ing v^{-\alpha } ,
\label{6}
\eeq
valid for $\Re v >0$ and $c\in \reals$ , $c>0$. Equation (\ref{6}) is 
easily proven by closing the contour to the left obtaining immediately
the power series expansion of $\exp (-v) $. 
Using (\ref{6}) in (\ref{5}) we find (defining $\mu_c =E_0$)
\bea
q&=& q_0 +\sum_{n=1}^{\infty} \frac 1 n e^{-n\beta (\mu_c -\mu)} 
{\sum_N}' e^{-n\beta (E_N -E_0)} \nn\\
&=&q_0 +\sum _{n=1}^{\infty}\frac 1 n e^{-n\beta (\mu_c -\mu)}\nn\\
&&\times\snp\ing 
(\beta n )^{-\alpha} \left(E_N -E_0 \right) ^{-\alpha}. \label{7}
\eea

At this stage we would like to interchange the summations over $N$ and $n$ and the integration in order to arrive at an expression containing the zeta function associated with the Schr\"odinger equation (\ref{1}),
\beq
\zeta (s) = \snp \en ^{-s}.\label{8}
\eeq
General zeta function theory tells us that the rightmost pole of $\zeta (s)$ 
is located at $s=d/2$, see for example \cite{voros}, 
followed by poles at $s=(d-1)/2,\ldots,1/2$; $-(2l+1)/2$, $l\in \nats_0$.
In general the 
rightmost pole appears at
$s=d/m$ where $d$ is the dimension of space and $m$ is the order of the 
elliptic differential operator, here two. Furthermore we need the polylogarithm,
\beq
Li_n (x) =\sum_{l=1}^{\infty} \frac{x^l}{l^n} ,\label{9}
\eeq
basic properties of which may be found in \cite{lewin,robinson}. 
As we will see in the following, the treatment of arbitrary dimension $d$ 
creates no additional complications. 
But it may not only be of academic interest. In this context we mention the 
analogies between bosons in a two-dimensional box and bosons in a 
one-dimensional harmonic oscillator potential analysed in Ref.~\cite{gert}. 
Similar analogies between higher dimensional cavities 
and external potentials in dimensions $d=1,2,3$ remain to be 
explored.

Due to the above remarks, in order that the summation and integration might
be interchanged we have to impose that $\Re c >d/2$ to obtain
\bea
q&=&q_0 +\ing \beta ^{-\alpha }\nn\\
&&\times Li_{1+\alpha} \left(e^{-\beta (\mu_c -\mu)}\right)
 \zeta (\alpha) .\label{10}
\eea
This is a very suitable starting point for the analysis of certain 
properties of the partition function $q$. Closing the contour to the
right corresponds to the large-$\beta$ expansion; closing it to the left to the small-$\beta$ expansion. (The relevant dimensionless expansion parameter will be made clear later). To the right of the contour the integrand in (\ref{10}) has no poles, which
 means that the large-$\beta$ behaviour contains no inverse power in $\beta$. One might show however, that the contribution from the contour itself is not vanishing at infinity leading to exponentially damped contributions for $\beta \to \infty$, the well
 known behaviour of partition sums at low temperature.

However, the relevant range for the analysis of BEC is the small-$\beta$ range.
(As we will see we are going to obtain a reliable expansion if the de
Broglie wavelength is small compared with the typical extensions of the
cavity. For $\mu m$-sized cavities this will be true near the transition
temperature.) Thus we close the contour to the left and pick
up only the leading two terms at $\alpha = d/2$ and 
(for $d>1$) $\alpha =(d-1)/2$. We find
\bea
q&=&q_0 +\Gamma (d/2) \beta^{-d/2} 
Li_{(d+2)/2}\muc 
\rez (d/2) \nn\\
& &+\Gamma \left(\frac{d-1} 2\right)  
\beta ^{-(d-1)/2} 
Li_{(d+1)/2} \muc\nn\\ 
&&\times\rez ((d-1)/2) +\ldots,\label{11}
\eea
with the residues of $\zeta (s)$ denoted by $\rez $.
The case $d=1$ is special in that $\zeta (s) $ has no pole at 
$s= (d-1)/2 =0$. For that dimension the subleading contribution contains
$\zeta ' (0)$ of which, as a rule, no detailed information is available. So
later, for $d=1$, we will consider only the leading contribution coming from
$\rez (1/2)$ and which is correctly given above.

In order to obtain a reliable approximation 
(in any dimension) at the critical temperature
where BEC occurs we are thus left with the task of determining the 
leading poles of the zeta function associated with the Schr\"odinger equation. Concerning this determination there are deep connections between the zeta function of an operator and its heat-kernel defined as \cite{gilkey}
\beq
K(t) = \snp e^{-t \en} .\label{12}
\eeq
The small-$t$ behaviour of (\ref{12}) is of special relevance here,
\beq
K(t) \sim \frac 1 {(4\pi t )^{d/2} } \sum_{l=0,1/2,1,\ldots}^{\infty} a_l t^l
,\label{13}
\eeq
where for a generally curved 
spacetime in the meantime the first $6$ coefficients are known
for an arbitrarily shaped smooth cavity \cite{gil1} and for 
an arbitrary second order elliptic operator with leading symbol the metric 
of the spacetime. We will need here only the first two coefficients which
for the Schr\"odinger operator at hand read
\bea
a_0 &=&\left(\frac{ 2m}{\hbar ^2} \right) ^{d/2} V;
\label{14a}\\
a_{1/2} &=& \left(\frac{ 2m}{\hbar ^2}\right)^{(d-1)/2}
\frac{\sqrt{\pi}} 2 (\pa V ) b,\label{14b}
\eea
with $V$ the volume of the cavity, $\pa V$ the area of its boundary, and
$b$ being a parameter depending on the boundary conditions,
\beq
b=\left\{ \begin{array}{rl}
 -1 & \mbox{for Dirichlet boundary conditions} \\
 1 & \mbox{for Neumann boundary conditions.}
\end{array}
\right.
\label{15}
\eeq
Although to the authors knowledge Neumann boundary conditions do not occur 
in physical applications of non-relativistic theories considered here, 
we include them for mathematical completeness and because they do not 
lead to any extra difficulty in calculation.
The relevance of Neumann boundary conditions in physics stems 
from the treatment of the electromagnetic field in the presence 
of ideal conducting plates, or the bag model in QCD. 

As is easily seen, 
the residues of $\zeta (s)$, equation (\ref{8})
might be expressed through the heat-kernel coefficients in equation
(\ref{13}). Using the integral representation for the $\Gamma$-function
we write
\beq
\zeta (s) = \frac 1 {\Gamma (s)} \snp \int_0^{\infty} dt \,\,
t^{s-1} e^{-\en }.\nn
\eeq
Then we split the integral into $t\in [0,1]$ and $t\in [1,\infty)$.
For the first interval we use the asymptotic expansion (\ref{13}), the
contribution from the
second interval is analytic in $s$ since the pole terms come from the $t\rightarrow0$ behaviour of the integrand. Thus we find
\bea
\zeta (s) &=& \frac 1 {(4\pi )^{d/2} } \frac 1 {\Gamma (s)} 
\sum_{l=0,1/2,1,\ldots} ^{\infty} a_l \int_0^1 dt\,\, 
t^{s-1-d/2+l}\nn\\
&&\quad + \mbox{ finite pieces }\nn\\
&=& \frac 1 {(4\pi )^{d/2} } \frac 1 {\Gamma (s)}
\sum_{l=0,1/2,1,\ldots} ^{\infty}
\frac{a_l}{s+l-d/2}\nn\\
&&\quad + \mbox{ finite pieces } \label{16}
\eea
and read off
\bea
\rez (d/2) &=& \frac{a_0}{(4\pi)^{d/2} \Gamma (d/2)};\label{17a}\\
\rez ((d-1)/2  ) &=& \frac{a_{1/2}}{(4\pi)^{d/2}\Gamma ((d-1)/2) }
. \label{17b}
\eea
Additional poles are located 
to the left of the above ones.
The associated residues will depend
on the details of the boundary of the cavity like its extrinsic 
curvature form. Having however nowadays experimental situations in mind,
these are probably negligible and so we are not going to present more than
the subleading order in our results. In doing this, results will be very 
comprehensible, but let us stress that higher orders could be included easily.

Using equations (\ref{17a},\ref{17b},\ref{14a},\ref{14b}) in the result (\ref{11}) for the
partition sum, we find to the two leading orders
(as mentioned, for $d=1$, now and in the following, only the leading piece 
is correct and will be considered)
\bea
q &=& q_0 +
Li_{(d+2)/2} \muc  
\frac V {\lambda _T ^d}\nn\\
&& +\frac{b} 4
\frac{\pa V}{\lambda _T ^{d-1}} Li_{(d+1)/2}\muc+\ldots,\label{18}
\eea
with the de Broglie wavelength 
\beq
\lambda_T = \frac{h}{\sqrt{2\pi m kT}}.\nn
\eeq
Here it is clearly seen, that the expansion parameter is given by 
(typical extension of the cavity)/$\lambda_T$. Our results are accurate for 
large values of the expansion parameter; this is, the expansion obtained is an expansion
about the thermodynamic limit, which is the relevant one for 
considering BEC. It is useful to introduce the dimensionless expansion 
parameter
\beq
\xi = \frac L {\lambda_T} \label{xi}
\eeq
with the definitions $V=L^d$ and $\partial V = \kappa L^{d-1}$. The dimensionless parameter $\kappa$ contains information related to the shape of the boundary. Then eq.~(\ref{18})
reads
\bea
q&=&q_0 +Li_{(d+2)/2} \muc \xi^d \nn\\
&&+\frac {\kappa b} 4 Li_{(d+1)/2} \muc 
\xi ^{d-1}+\ldots \label{18b}
\eea
At sufficiently low temperatures, where the ground state population is large,
the approximation $\mu \simeq \mu_c =E_0$ holds. Then using $Li_n (1) = \zeta
_R (n)$, $n>1$, eq.~(\ref{18}) simplifies to
\bea
q&=&q_0 +\zeta_R ((d+2)/2) \frac V {\lambda_T^d}\nn\\
&& +\frac b 4 \zeta_R ((d+1)/2) \frac{\partial V}{\lambda_T^{d-1}}+\ldots,\label{18a}
\eea
which is a very good approximation at some 
range below the critical temperature.

Let us now continue with the particle number,
\beq
N= \beta^{-1} \left(\frac{\pa q}{\pa \mu}\right) \left.\right|_{T,V}.\label{19}
\eeq
Here one has 
\beq
N=N_0 +\snp \frac{e^{-\beta \en }}{1-e^{-\beta \en }},\label{20}
\eeq
with the ground state occupation number
\beq
N_0 = \frac{d_0}{e^{\beta (\mu_c -\mu )}-1}.\label{21}
\eeq
(Recall that $\mu_c=E_0$.) The high temperature expansion of $N$ is easily obtained by noting 
\beq
\frac{\partial Li_n (x)}{\partial x} = \frac 1 x Li_{n-1} (x). \label{poly}
\eeq
It reads
\bea
N&=&N_0 +Li_{d/2} \muc \xi ^d \nn\\
&&+ \frac{\kappa b} 4 Li_{(d-1)/2} \muc \xi ^{d-1}
+\ldots\label{n23}
\eea
The critical temperature $T_c$ (in the absence of a phase transition) 
can be defined as the temperature at which the number of particles in the 
ground state begins to become large. We will set $N=fN_0$ with 
$f$ representing the fraction of particles in the
ground state. If we assume $N_0>>1$, then for temperatures close to 
$T_c$ we expect $\mu\simeq\mu_c$. (See Eq.~(\ref{21}).) 
The behaviour of the polylogarithms in (\ref{n23}) depends on the 
dimension $d$. For $d>3$ all of the terms in (\ref{n23}) are 
finite as $\mu\rightarrow\mu_c$, and we can approximate (\ref{n23}) by
\bea
N &\simeq& \zeta_R (d/2) \xi_c^d \nn\\
&&+\frac{\kappa b} 4 \zeta_R ((d-1)/2)
\xi_c^{d-1} +\ldots\;. \label{n25}
\eea
The critical temperature is defined via $\xi_c$ obtained by taking $T=T_c$ in (\ref{xi}). This assumes $f<<1$.
Defining 
\beq
\xi_0^d = \frac N {\zeta_R (d/2)} , \label{n27}
\eeq
which gives the critical temperature in the bulk (or thermodynamic) limit,
\beq
T_0 = \frac{h^2} {2\pi mk} \left( \frac N {V\zeta_R (d/2)} \right)
^{2/d}.\label{n29}
\eeq
and assuming that 
$\xi_c = \xi_0 (1+\gamma )$, with $\gamma \ll 1$, to leading order we can approximate (we use the temperature here) 
\beq
T_c = T_0 \left( 1-\frac{\kappa b}{2d} \frac{ \zeta_R ((d-1)/2)}{ \zeta_R
^{(d-1)/d} (d/2) } \frac 1 {N^{1/d}}+\ldots\right) . \label{n28}
\eeq
It is seen, that depending on the boundary conditions imposed, the 
critical temperature can increase or decrease. The magnitude of the 
correction behaves like $N^{-1/d}$ and is typically not negligible. It is seen
explicitly that the corrections are going to vanish for $N\to \infty$.  

Let us now consider the lower dimensional cases and we start with $d=3$. 
Then we need 
\beq
Li_1 \muc = -\ln (1-\muc). \label{au1}
\eeq
Near $\mu \simeq \mu_c$ we use eq.~(\ref{21}) to approximate
\beq
\beta (\mu _c -\mu ) =\ln \left( 1 + \frac 1 {N_0} \right) \simeq 
\frac 1 {N_0}. \label{au2}
\eeq
Then in 3 dimensions the analogue of eq.~(\ref{n25}) reads
\beq
N(1-f) \simeq \zeta_R (3/2) \xi ^3 + \frac{\kappa b } 4 \ln (fN) \xi ^2+\ldots
\label{au3}
\eeq
Here it is not possible to put $f=0$ in order to define a critical
temperature, since in order to obtain (\ref{au3}) we have 
assumed $N_0>>1$. (See Eq.~(\ref{au2}).) This roughly 
reflects the fact that in 3 dimensions (as well as in 1 and 2 as seen below) 
the number of excited states is 
lower than in higher dimensions, 
and as a result, in the temperature range considered, 
particles have to reside in the ground state in order that the 
thermodynamic equation for $N$ is fulfilled. However given 
$f\neq 0$ and $N$, this might be solved for the 
temperature,
\beq
T=T_0 \left\{ 1-\frac 2 3 (f+\alpha ) \right\} \label{au4}
\eeq
with 
\beq
T_0 = \frac{ h^2 } {2\pi mk} \left( \frac N {\zeta_R (3/2) V } \right) 
^{2/3} \nn
\eeq
and 
\beq
\alpha = \frac{ \kappa b \ln (fN) }{ 4 \zeta_R ^{2/3} (3/2) N^{1/3} },
\nn
\eeq
valid near $\mu \simeq \mu_c$. This parallels completely eq.~(\ref{n28}).
Defining the critical temperature $T_c$ as the temperature where a given  
fraction $f$ of the number of particles $N$ is in the ground state,
eq.~(\ref{au4}) can easily be used to determine $T_c$. 
This equation also shows that in the thermodynamic limit, 
$f$ may be put to zero and $T_0$ is a "real'' critical temperature. 
This reflects the fact that in three-dimensional 
unbounded space the ideal
Bose gas exhibits Bose-Einstein condensation in the sense of a phase transition.

We proceed with $d=2$ and need in addition 
\beq
Li_{1/2} \muc \simeq \sqrt{\frac{\pi}{\beta (\mu_c -\mu ) }  } +\ldots
\label{au5}
\eeq
valid for $\mu \simeq \mu_c$. Using the same approximation as before,
$\beta (\mu_c -\mu ) \simeq 1/N_0$, this results in 
\beq
N(1-f) =\xi^2 \ln (fN) +\frac{\kappa b \sqrt{\pi} } 4 \xi \sqrt{fN} +\ldots
\label{au6} 
\eeq
Solving to leading order for $\xi$ given $f\simeq 0$ we find
\beq
\xi _c =\sqrt{\frac N {\ln (fN) } } \label{au7} 
\eeq
or in the temperature
\beq
T _c= \frac{h^2} {2\pi km} \frac N {\ln (fN) V } . \label{au8}
\eeq
Finally in $d=1$ considering only the leading contribution results in
\beq
\xi _c= \sqrt{ \frac N {\pi f}  } \label{au9} 
\eeq
with
\beq
T _c= \frac{h^2} { 2\pi km } \frac N {\pi fV} .\label{au}
\eeq
In dimensions $d=1$ and $d=2$ it is clearly seen that we cannot put $f=0$. This reflects the fact that even in the thermodynamic limit Bose-Einstein condensation (as a phase transition) is not possible.  

Another interesting quantity of the system is its internal energy defined by
\beq
U =\left\{-\frac{\pa}{\pa \beta} +\frac{\mu}{\beta} \frac{\pa}{\pa\mu}
\right\} q.\label{28}
\eeq
Of course the energy of the system is the sum of the single particle energies $E_N$ multiplied by the occupation number,
\beq
U=U_0 +\snp E_N \frac{e^{-\beta \en }}{1-e^{-\beta \en }}.\label{29}
\eeq
Here $U_0$ is the zero-point energy,
\beq
U_0 = \frac{d_0 E_0}{e^{\beta (E_0 -\mu )}-1}.\label{29a}
\eeq
To the approximation considered, $U$ is most easily found by using 
(\ref{11}) and (\ref{poly}). As an immediate result we have
\bea
U&=& -\frac{\partial q} {\partial \beta} +\mu N \nn\\
 &=& U_0 +\frac d 2 \beta^{-1} Li_{(d+2)/2} \muc \xi ^d \nn\\
& &+\frac{\kappa b} 8 (d-1) \beta^{-1} Li_{(d+1)/2} \muc \xi ^{d-1} \nn\\
& &+E_0 Li_{d/2} \muc \xi ^d \label{au11}\\
& &+E_0 \frac{\kappa b} 4 Li_{(d-1)/2} \muc \xi ^{d-1} +\ldots\nn
\eea
With the already presented expansions around $\mu \simeq \mu_c$ the behaviour
near the phase transition is easily displayed.

Finally we are left with considering the specific heat,
\beq
C =\left( \frac{\pa U}{\pa T}\right) _{N,V}. \label{32}
\eeq
The slightly more difficult point here is, that $N$ has to be considered as
fixed and so $\mu$ has to be considered as a function of $N$ and $\beta$. We 
will need $(\pa/\pa \beta) (\beta \mu) $, which is found by differentiating
equation (\ref{20}) given that $N$ is fixed, 
\beq
\left.\frac{\pa N}{\pa \beta} \right|_{N,V} =0 .\label{32a}
\eeq
Using eq.~(\ref{n23}) to the leading order, condition (\ref{32a}) yields
\bea
\frac{\partial}{\partial \beta } \left[
\beta (E_0 -\mu)\right] =\hspace{4.5cm}&&\nn\\ 
-\frac{ (d/2) \beta^{-1} Li_{d/2}   
\muc \xi ^d } 
{d_0 \frac \muc {(1-\muc)^2} +Li_{d/2 -1} \muc \xi^d }.&&
\label{implicit}
\eea
It is seen, that in the approximation $\mu = \mu_c$ one finds the above 
equation to yield zero as it should. 
Neglecting pieces coming from eq.~(\ref{implicit}) 
(one can show that this is justified for $d<7$) we only have 
to use (\ref{au11}) to find
\bea
\frac C k &=& \frac d 2 \left( 1+\frac d 2 \right) Li_{(d+2)/2} 
\muc \xi ^d \nn\\
& &+\frac{\kappa b} 8  \left( \frac{d^2-1} 2 \right) 
Li_{(d+1)/2} \muc \xi^{d-1}\nn\\
&& +\ldots\label{specific}
\eea
showing once more the decrease or increase in the specific heat depending 
on the 
boundary conditions.
Using previous formulas, expansions for $\mu \simeq \mu_c$ and for 
specific dimensions are easily found.

A slightly different treatment, which we do not pursue in this paper, could also be used. This uses the effective
fugacity (for the anisotropic harmonic oscillator see \cite{ravndal})
\beq
z_{eff} = z e^{-\beta E_1} \nn
\eeq
where $E_1$ is the first excited level with degeneracy $d_1$.
Whereas in the calculations presented above only the ground state has been treated
separately. We can now separate the ground state and the first excited level
to find
\bea
q& =& q_0 +d_1 Li_1 (z_{eff}) + 
       Li_{(d+2)/2} (z_{eff}) \frac V {\lambda_T^d}\nn\\
&& + 
\frac b 4 \frac{\pa V}{\lambda_T ^{d-1}} Li_{(d+1)/2} (z_{eff}) + \ldots 
\nn
\eea
This expansion is a very good approximation
even below the critical temperature
(for the harmonic oscillator see \cite{ravndal}).
All thermodynamical properties are obtained in the same manner as
before.

\section{Density of states method}

An alternative approach to the use of the Mellin-Barnes contour integral 
representation is to try to convert the sums for the thermodynamic 
quantities directly into ordinary integrals with an appropriate density 
of states factor. This has been widely used in a variety of problems 
in statistical mechanics. (~See Ref.~\cite{PathriaSM} for example.) 
Recently Grossmann and Holthaus \cite{GHPLA,GHNat} have used such a 
density of states method to study the statistical mechanics of particles 
in a harmonic oscillator potential trap, and in a 3-dimensional cubic 
box \cite{10} with Dirichlet boundary conditions. The harmonic 
oscillator trap is characterized by a density of states
\beq
\rho(E)=\frac{E^2}{2\Omega^3}+\gamma\frac{E}{\Omega^2}\;,\label{3.1}
\eeq
where $\Omega=(\omega_1\omega_2\omega_3)^{1/3}$ is the geometric mean of 
the anisotropic harmonic oscillator frequencies. 
In Ref.~\cite{KKDJTPLA} 
we showed how it is possible to evaluate $\gamma$ analytically, 
and obtain additional corrections to (\ref{3.1}). (See also \cite{ravndal}.) The purpose of the 
present section is to discuss the method in greater detail in a more 
general setting, and to show how results equivalent to those of the 
previous section may be obtained with the density of states method. 
The basic idea behind the method can be found in Ref.~\cite{BaltesHilf} 
with a number of examples illustrated.

Suppose that we have a self-adjoint differential operator $\Delta$ 
which possesses a set of eigenvalues $\lbrace\lambda_N\rbrace$ which 
are all positive. Here $N$ represents a multi-index which labels the 
eigenvalues. For the case of statistical mechanics in a cavity, 
$\Delta$ can be taken to be the Hamiltonian with $\lambda_N$ the 
energy levels $E_N$. The eigenvalues depend on the details of the cavity 
and boundary conditions imposed, but the method we describe can be 
used regardless of such details. Let ${\cal N}(\lambda)$ be the number 
of modes for the boundary value problem in the cavity for which the 
eigenvalues $\lambda_N\le\lambda$. We can write
\beq
{\cal N}(\lambda)=\sum_N\theta(\lambda-\lambda_N)\label{3.2}
\eeq
where $\theta(x)$ is the Heaviside distribution (or step function) defined by
\beq
\theta(x)=\left\lbrace\begin{array}{cc} 1,&\ x>0\\
	0,&\ x<0\\
	\frac{1}{2},&\ x=0\end{array}\right.\;\;.\label{3.3}
\eeq
This gives the result
\beq
{\cal N}(\lambda)=\sum_{\lambda_N<\lambda}1+\sum_{\lambda_N=\lambda}
\frac{1}{2}\;,\label{3.4}
\eeq
as used by Baltes and Hilf~\cite{BaltesHilf}.

The aim now is to treat the eigenvalues as a continuous distribution 
rather than a discrete set by introducing the eigenvalue (or spectral) 
density $\rho(\lambda)$. We define
\bea
\rho(\lambda)d\lambda&=&{\cal N}(\lambda+d\lambda)-{\cal N}(\lambda)
\label{3.5}\\
&\simeq&\frac{d{\cal N}(\lambda)}{d\lambda}d\lambda\;\;.\label{3.6}
\eea
Use of (\ref{3.2}) for ${\cal N}(\lambda)$, along with the 
distributional identity $\theta'(x)=\delta(x)$ gives
\beq
\rho(\lambda)=\sum_N\delta(\lambda-\lambda_N)\;.\label{3.7}
\eeq
If we use the familiar exponential representation for the Dirac delta 
distribution, then
\bea
\rho(\lambda)&=&\sum_N\frac{1}{2\pi}\int_{-\infty}^{\infty}
dk\,e^{ik(\lambda-\lambda_N)}\nn\\
&=&\sum_N\frac{1}{2\pi i}
\int_{-i\infty}^{i\infty}dt\,e^{t(\lambda-\lambda_N)}\;.\label{3.8}
\eea
We may now translate the integration contour in (\ref{3.8}) to the 
right by an amount $c$ where $c\in\reals$ with $c>0$, and assume that 
we are permitted to interchange the order of the summation and 
integration. This results in
\beq
\rho(\lambda)=\frac{1}{2\pi i}
\int_{c-i\infty}^{c+i\infty}dt\,e^{t\lambda}K(t)\;,\label{3.9}
\eeq
where
\beq
K(t)=\sum_Ne^{-t\lambda_N}\;.\label{3.10}
\eeq
Eq.~(\ref{3.9}) shows that $\rho(\lambda)$ is the inverse Laplace 
transform of $K(t)$. Inversion of (\ref{3.9}) results in
\beq
K(t)=\int_{0}^{\infty}d\lambda\,e^{-t\lambda}\rho(\lambda)\;.\label{3.11}
\eeq
Another way of arriving at (\ref{3.9}) is to note that (\ref{3.11}) 
follows directly from (\ref{3.7}) and (\ref{3.10}), establishing that 
$\rho(\lambda)$ is the Laplace transform of $K(t)$. The inversion formula 
for Laplace transforms gives (\ref{3.9}) immediately without the 
intermediate steps in (\ref{3.8}).

The point of (\ref{3.9}) and (\ref{3.11}) is that given $K(t)$ as defined 
in (\ref{3.10}) we can evaluate the density of states in a 
straightforward way. In certain simple cases, such as the harmonic 
oscillator potential, where the eigenvalues can be written down 
explicitly, it is possible to perform the sum in (\ref{3.10}) and 
evaluate $K(t)$ in closed form. (See Ref.~\cite{KKDJTPLA} for a case of 
this.) In the generic case, where the eigenvalues are not known explicitly, 
we can still obtain a result for the density of states by using the known 
asymptotic expansion for $K(t)$ as $t\rightarrow0$ \cite{gilkey}.

Assume that as $t\rightarrow0$ we have the generic expansion
\beq
K(t)\simeq\sum_{i=1}^{k}c_it^{-r_i}+{\cal O}(t^{-r_{k+1}})\label{3.12}
\eeq
for some coefficients $c_i$ and powers $r_i$ with $r_1>r_2>\cdots >r_k$. 
The simplest way to evaluate $\rho(\lambda)$ is to use (\ref{3.11}) 
and the relation
\beq
t^{-z}=\frac{1}{\Gamma(z)}\int_{0}^{\infty}d\lambda\,\lambda^{z-1}
e^{-t\lambda}\label{3.13}
\eeq
 which is valid for $\Re z>0$. It is now easy to see that
\beq
\rho(\lambda)\simeq\sum_{i=1}^{k}
\frac{c_i}{\Gamma(r_i)}\lambda^{r_i-1}\;.\label{3.14}
\eeq
Integrating this result with respect to $\lambda$ results in
\beq
{\cal N}(\lambda)\simeq\sum_{i=1}^{k}
\frac{c_i}{\Gamma(r_i+1)}\lambda^{r_i}\;.\label{3.15}
\eeq

Although the analysis we have presented makes no attempt at proper 
mathematical rigour, the main result in (\ref{3.14}) of (\ref{3.15}) is 
in agreement with a refinement of Karamata's theorem due to Brownell 
\cite{Brownell}. (A nice account is contained in Ref.~\cite{BaltesHilf}.) 
It is important to emphasize that the result for 
$\rho(\lambda)$ or ${\cal N}(\lambda)$ refers really to the average number 
of eigenvalues. Generally the number will fluctuate around some average 
value. These fluctuations can be important in some contexts 
\cite{BaltesHilf,BerryHowls}. Another point worth stressing is that our 
results have assumed that $r_k$ in (\ref{3.12}) is positive. The results 
(\ref{3.14}) and (\ref{3.15}) are actually true for all $r_k$ if we 
define $1/\Gamma(r_i)=0$ when $r_i=0,-1,-2,\ldots$. The establishment 
of this result requires a more powerful approach than 
Laplace transforms \cite{Brownell}.

The density of states (\ref{3.14}) is determined by a knowledge of the 
coefficients $c_i$ appearing in the asymptotic expansion of $K(t)$. 
These coefficients can be evaluated from a knowledge of the generalized 
$\zeta$-function associated with the eigenvalues $\lambda_N$ as discussed 
in the previous section. Defining
\beq
\zeta(s)=\sum_N\lambda_N^{-s}\label{3.16}
\eeq
the analysis in Sec.~2 is easily modified to show that $\zeta(s)$ has a 
simple pole at $s=r_i$ with residue
\beq
\left.{\rm Res}\,\zeta(s)\right|_{s=r_i}=\frac{c_i}{\Gamma(r_i)}\label{3.17}
\eeq
for $r_i\ne0,-1,-2,\ldots$; and
\beq
\zeta(s=-r_i)=(-1)^{r_i}(-r_i)!c_i\label{3.18}
\eeq
for $r_i=0,-1,-2,\ldots$. Therefore the coefficients $c_i$ appearing in 
the density of states may also be found from a knowledge of the generalized 
$\zeta$-function. This provides a direct link between the density of 
states approach and the method described in Sec.~2.

We will conclude this section by showing how the density of states 
method may be used in the specific case of an ideal gas confined in a 
general cavity. The $q$-potential defined earlier was written as
\beq
q=q_0+q_{ex}\label{3.19}
\eeq
where
\beq
q_0=- d_0 \ln\Big(1-ze^{-\beta E_0}\Big)\label{3.20}
\eeq
was the ground state contribution, and 
\beq
q_{ex}=\sum_{n=1}^{\infty}
\frac{e^{n\beta(\mu-\mu_c)}}{n}\snp e^{-n\beta(E_N-E_0)}\label{3.22}
\eeq
represents the contribution from the excited states.
The sum over energy levels in (\ref{3.22}) can now be converted into an 
integral over a continuous energy variable by introducing a density of 
states factor. From (\ref{3.14}) and using the coefficients $c_i$ 
found in Sec.~2 we have
\beq
\rho(E)\simeq
\frac {V\beta^{d/2}E^{d/2-1} }{\lambda_T^d \Gamma (d/2)} +
\frac b 4 \frac{(\partial V) \beta^{(d-1)/2} E^{(d-3)/2} } 
{\lambda_T^{d-1} \Gamma ((d-1)/2) }    
\label{3.23}
\eeq
if we keep only the first order correction to the bulk expression as 
before. To this order we may approximate
\beq
q_{ex}\simeq\sum_{n=1}^{\infty}\frac{e^{n\beta(\mu-\mu_c)}}{n}
\int_{E_1-E_0}^{\infty}dE\,\rho(E)e^{-n\beta E}\;.\label{3.24}
\eeq
Replacing 
the lower limit in the integration with 0, we obtain
\bea
q_{ex}&\simeq&\lambda_T^{-d}V\,Li_{(d+2)/2}\left(e^{-\beta(\mu_c-\mu)}\right)\nn\\
&&+\frac{b}{4}\lambda_T^{-(d-1)}(\partial V)
Li_{(d+1)/2}\left(e^{-\beta(\mu_c-\mu)}\right)\label{3.25}
\eea
in agreement with our earlier method. The replacement of the lower 
limit on the integral with 0 assumes $\beta E_0<<1$ so that the 
temperature is well above the temperature associated with the ground 
state energy ({\em i.e.} $kT>>E_0$.) If the energy gap $E_1-E_0$ is 
of the same order of magnitude as the ground state energy, then 
$\beta E_1<<1$ as well. Results for the particle number and other 
thermodynamic quantities follow as before.

\section{Application of the results to some cavities}

This section will discuss some specific examples like the rectangular box
and the spherical cavity 
with various boundary conditions. The rectangular box has been the subject of considerable investigation. (See Refs.~\cite{Kac,PathriaCJP} for reviews.) In order to get all 
thermodynamic quantities as derived in Secs.~II and III, we only need to
give the constant $b$, the volume of the cavity $V$, and the boundary $\partial V$. Another interesting quantity is the lowest
energy eigenvalue as the critical value for the chemical potential.
\subsection{Rectangular box}
For the rectangular box with side lengths $L_1,\ldots,L_d$ one can consider
for example periodic, Dirichlet and Neumann boundary conditions. In 
cartesian coordinates $x=(x_1,\ldots,x_d)$ with $x_i \in [0,L_i]$.

For periodic boundary conditions, 
\beq
\phi (x_1,\ldots,x_i,\ldots,x_d) =
\phi (x_1,\ldots,x_i+L_i,\ldots,x_d),
\nn
\eeq 
with $i=1,\ldots,d$, the energy eigenvalues are
\beq 
E_{n_1,\ldots,n_d} = \sum_{i=1}^d \left(\frac{2\pi} {L_i} n_i \right) ^2 ,
\qquad n_i \in \intgs \nn.
\eeq
Periodic identification means, that effectively there is no boundary. (Equivalently we can regard $V$ as the $d$-torus.) In this case $b=0$. Furthermore, one has for the volume
\beq
V_d = \prod_{i=1}^d L_i \nn
\eeq
and $\mu_c=E_0=0$.

For Dirichlet boundary conditions, 
\bea
\phi (x_1,\ldots,x_{i-1},0,x_{i+1},\ldots,x_d)&=&0,\nn\\
\phi (x_1,\ldots,x_{i-1},L_i,x_{i+1},\ldots,x_d)&=&0,\nn
\eea
one gets
\beq
E_{n_1,\ldots,n_d} = \sum_{i=1}^d \left( \frac{\pi n_i} {L_i} \right) ^2,
\qquad n_i \in \nats . \nn
\eeq
As mentioned, see eq.~(\ref{15}), $b=-1$, the volume is as before and 
furthermore
\beq
\partial V = 2 \sum_{i=1}^{d} V_{d-1}^{(i)}.\nn
\eeq
Here, $V_{d-1}^{(i)}$ is the volume of the i'th side of the rectangular 
cavity, $V_{d-1}^{(i)} =L_1\ldots L_{i-1}L_{i+1}\ldots L_d$. Furthermore,
\beq
E_0 = \sum_{i=1}^d \left(\frac{\pi}{L_i}\right)^2.\nn
\eeq
If we specialize to the $d$-cube, then $\partial V=2dL^{d-1}$ which gives $\kappa=2d$.

Finally, for Neumann boundary conditions,
\bea
(\partial /\partial x_i ) \phi (x_1,\ldots,x_d) |_{x_i =0} &=&0\nn\\
(\partial /\partial x_i ) \phi (x_1,\ldots,x_d) |_{x_i =L_i}&=&0\nn ,
\eea
the energy eigenvalues read
\beq
E_{n_1,\ldots,n_d} = \sum_{i=1}^d \left( \frac{\pi n_i}{L_i}\right)^2 ,
\qquad n_i \in \nats_0,\nn
\eeq
with $\mu_c=E_0=0$, $b=1$ and $V$, $\partial V$, as before.

All thermodynamical quantities may now be given immediately with the 
formulas in Sec.~II. Especially for the critical temperature one has 
that periodic boundary conditions give a critical temperature identical to 
the thermodynamic limit (a result of being no boundary there), Dirichlet
boundary conditions increase it and Neumann boundary conditions decrease it.

Looking at the distribution of the eigenvalues around the ground state of 
the system an intuitive explanation of this behaviour is possible. Let us
assume for this equal compactification sides $L_i$. It is clear, the lower
the density of eigenvalues near the ground state, the higher the transition
temperature will be. Then the following might be stated. Neumann boundary
conditions lead to the lowest condensation temperature because it has the
smallest spacing between the ground state and the first excited level.
This compensates the doubled number of eigenstates in the periodic case,
the eigenvalues which are however four times higher. The difference between
Neumann and Dirichlet boundary conditions is based on the same observation; in addition the degeneracy of the eigenvalues is the same for this two 
boundary conditions. The difference between periodic and Dirichlet 
boundary conditions is not obvious, however the results show, that the
slightly smaller level spacing for Dirichlet boundary conditions is more than
compensated by the doubling of the degeneracy for periodic boundary conditions.

\subsection{Spherical cavities}
For spherical cavities an explicit knowledge of the eigenvalues
is not at hand and for that reason also a numerical treatment of
the partition sum is not immediate. For this kind of examples our
analytical approach is most useful. 

It is convenient to introduce a spherical coordinate basis, with $r=|x|$
and $d-1$
angles $\Omega =(\theta_1 ,\ldots,\theta_{d-2},\varphi )$.
In these coordinates, a complete set of solutions of the Schr\"odinger
equation  
together with one of the mentioned boundary conditions may be given in
the form
\beq
\phi_{l,m,n}(r,\Omega )=r^{1-\frac d 2} J_{l+\frac{d-2}2} (\sqrt{2m}w_{l,n}r
/\hbar)
Y_{l+\frac d 2} (\Omega ),
\eeq
with $J_{l+(d-2)/2}$ being Bessel functions and
$Y_{l+d/2}$ hyperspherical harmonics
\cite{erdelyimagnusoberhettingertricomi53}. 
Here, $E_{l,n}=w^2_{l,n}$ and the $w_{l,n}
(>0)$ are determined through the boundary conditions by
\beq
J_{l+\frac{d-2} 2 }(\sqrt{2m}w_{l,n}R/\hbar)
= 0,
\eeq 
for Dirichlet boundary
conditions and 
\bea
0&=&\frac {1-d/2} R J_{l+\frac{d-2} 2 }(\sqrt{2m}w_{l,n}R/\hbar)\nn\\
&&+
\frac{\sqrt{2m}w_{l,n}}{\hbar} J'_{l+\frac{d-2} 2 }(\sqrt{2m}
w_{l,n}r/\hbar)\left|_{r=R} \right.\nn
\eea
for Neumann
boundary conditions.
Although none of these eigenvalues is known explicitly, in order to obtain 
an approximation of all thermodynamic quantities, it is already enough 
to state the volume and the boundary volume of the spherical cavity,
\beq
V=\frac{\pi^{d/2} R^d }{ \Gamma (d/2+1)}\;\;,
\partial V =\frac{2\pi^{d/2} R^{d-1} }{\Gamma (d/2)}, \nn
\eeq
and to apply the results of the previous sections.

\section{Discussion and conclusions}

We have described two methods for treating the ideal Bose gas using the grand canonical ensemble in a general cavity. The first method, described in Sec.~II, treats the partition function as a sum over the discrete energy levels by making use of the 
Mellin-Barnes
integral representation for the exponential function. The second method 
shows how to obtain an adequate density of states factor 
so that the sum over discrete energy levels can be replaced with an integral. 
Both methods make use of the so-called heat-kernel coefficients for the 
Schr\"{o}dinger operator, and equivalence between
the two methods is shown in quite a general setting.

Our analysis shows clearly the way in which the geometry of the 
cavity enters in all thermodynamic properties. In the limit 
considered, namely $\xi=L/\lambda_T\gg1$, the leading 
correction to the bulk limit term $\xi^d$ enters via the property 
$\kappa=(
\partial V)/V^{(d-1)/d}$. 
Thus the leading order correction is completely described by the 
volume and area of the cavity, with all finer details washed out 
for $\xi\gg1$. Finer detail would be present in the next to 
subleading order in terms of the extrinsic
curvature of the boundary. 
(The extrinsic curvature describes how the normal vector to the boundary varies when moving along the boundary.)

These ideas are as well applicable to the canonical ensemble. Furthermore,
due to the connections between the microcanonical and the grand canonical
approach stated recently in \cite{mikro1,neu1,neu2} it seems possible to develop
this approach also for the microcanonical ensemble.

\acknowledgements
This investigation has been partly supported by the
DFG under contract number
BO1112/4-2.

\end{document}